\documentclass{iopart}

\usepackage{epsfig}

\begin{document}

\title{Hadronization of Dense Partonic Matter}

\author{Rainer J Fries}
\address{School of Physics and Astronomy, University of Minnesota, \\
Minneapolis, MN 55455}
\ead{fries@physics.umn.edu}

\begin{abstract}
The parton recombination model has turned out to be a valuable tool
to describe hadronization in high energy heavy ion collisions. 
I review the model and revisit recent progress in our understanding of 
hadron correlations. I also discuss higher Fock states in the hadrons, 
possible violations of the elliptic flow scaling and recombination effects 
in more dilute systems.
\end{abstract}

\submitto{JPG}

\section{Introduction}

Hadronization of partons is one of the unsolved problems in Quantum
Chromodynamics (QCD). Hadronization is a non-perturbative process and
deeply connected to confinement which by itself is still not understood
on a fundamental level.
Fortunately, techniques have been invented that help to deal with
hadronization phenomena in elementary particle collisions in which hadrons
are produced. They are based on universal parameterizations of the unknown
non-perturbative quantities. A famous example are fragmentation functions 
$D_{a/A} \sim \langle 0|u|h(P)\rangle\langle h(P)|u|0\rangle$ which
parameterize the probability that a hadron $h$ with large momentum $P$ is 
created from a parton $u$ with momentum $p>P$ in the vacuum \cite{CoSo:81}.
It has been shown that for hadron production at asymptotically large 
momentum transfer the cross section factorizes into a fragmentation
function and a perturbative cross section between partons in a well-defined 
way. 
A second example are exclusive processes where the wave functions 
$\psi \sim \langle 0| u u d|h(P) \rangle$ of the 
(few) lowest Fock states of a hadron appear \cite{CZER:77}.

Yet another method to deal with hadronization in a simplified way is the
quark recombination model \cite{DasHwa:77}. As in exclusive hadron production 
the matrix elements tested are of the same form as the one in exclusive
processes $\sim \langle 0| u u d|h\rangle$. This is only a formal 
correspondence, but attempts have been made to connect recombination to 
both fragmentation \cite{MiJoLa:81} and to exclusive hadron production
\cite{Ochiai:85}.

Recombination can be directly applied to describe hadron production
in rather dense parton systems \cite{FMNB:03prl,FMNB:03prc,GreKoLe:03prl,
HwaYa:02,Fries:04qm, Fries:04hq}. The motivation is twofold. First, 
there is an expectation that
in a dense parton phase fragmentation is not the correct picture for hadron
formation. Secondly, it became indeed clear that perturbative hadron 
production and fragmentation are not sufficient to explain RHIC data for 
hadrons at transverse momenta of several GeV/$c$. This conclusion is mainly 
based on the observed large baryon/meson ratio \cite{PHENIX:03ppi}, the 
nuclear modification factors $R_{AA}$ (or $R_{CP}$) close to or above unity 
\cite{PHENIX:03ppi,STAR:03llbar,STAR:03v2} and the remarkable scaling law 
for elliptic flow \cite{PHENIX:03v2,STAR:03v2}.

\section{Hadronization of Bulk Matter}

In central heavy ion collisions a hot and dense fireball of deconfined quarks 
and gluons is created.
In the recombination model one postulates the existence of thermalized parton 
degrees of freedom at the phase transition temperature $T_c$ which 
recombine or coalesce into hadrons. It has been found to be sufficient 
to consider the lowest Fock state in each hadron, the valence quarks, 
which are given constituent masses around 300 MeV.


The spectrum of hadrons can be calculated starting
from a convolution of Wigner functions \cite{FMNB:03prc}. 
For a meson with valence (anti)quarks $a$ and $b$ we have
\begin{equation}
  \label{eq:reco}
  \fl
  \frac{d^3 N_M}{d^3 P}= \sum_{a,b} \int\frac{d^3 R}{(2\pi)^3} 
  \int\frac{d^3 qd^3 r}{(2\pi)^3} W_{ab}\left(\mathbf{R}-
  \frac{\mathbf{r}}{2},\frac{\mathbf{P}}{2}-\mathbf{q}; \mathbf{R}+
  \frac{\mathbf{r}}{2},\frac{\mathbf{P}}{2}+\mathbf{q} \right)
  \Phi_M (\mathbf{r},\mathbf{q}).
\end{equation}
Here $W_{ab}$ is the 2-particle Wigner function for partons $a$, $b$ and 
$\Phi_M$ is the Wigner function of the meson. The sum runs over all
possible parton quantum numbers. For simplicity
the parton Wigner function is usually approximated by a product
of single particle phase space distributions $W_{ab}=w_a w_b$.
Several slightly different implementations of this formalism have been 
discussed in the literature \cite{FMNB:03prc,GreKoLe:03prl,HwaYa:02}.
See \cite{Fries:04qm,Fries:04ca} for earlier reviews.

In order to obtain an estimate when recombination is important as a
hadronization mechanism, one can compare the yields of fragmentation 
and recombination starting from different parton spectra. Thermal
parton spectra $w\sim e^{-P/T}$ play a special role. Recombination of thermal
partons leads to an exponential hadron distribution 
with the same slope since
\begin{equation}
  w_a w_b \sim \sim e^{-xP/T} e^{-(1-x)P/T} = e^{-P/T}.
\end{equation}
where $x$ gives the momentum fraction of parton $a$. 
Therefore recombination is more effective than fragmentation on any 
thermalized parton ensemble. On the other hand one can show that a power-law
parton spectrum favors fragmentation at least for large $P_T$, in accordance
with perturbative QCD.

Eq.\ (\ref{eq:reco}) does preserve momentum, but not energy. Therefore it
can only be safely applied in a kinematic region where mass effects are
small, i.e.\ for $P_T \gg M$. On the other hand fragmentation starts to 
dominate at very high $P_T$. The large jet quenching
at RHIC further suppresses the contribution from fragmentation, so that 
recombination effects can be observed at intermediate $P_T$.

\begin{figure}
\begin{center}
\epsfig{file=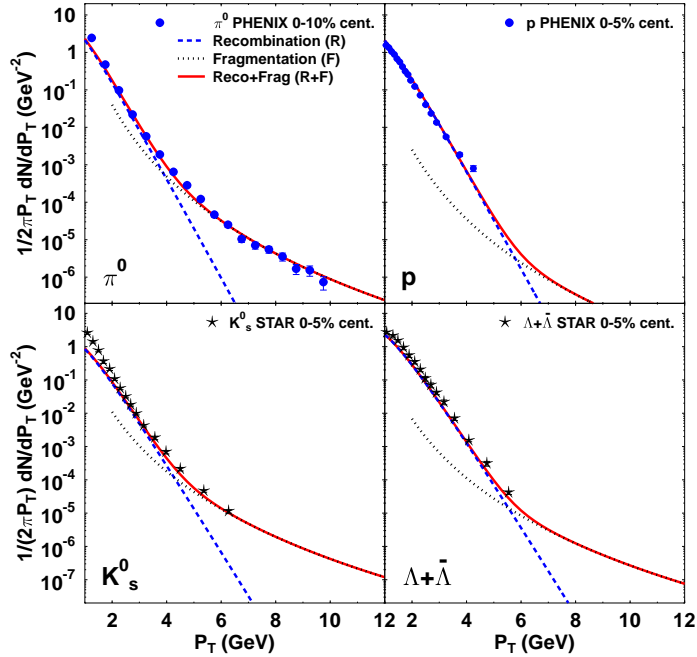,width=10cm}
\caption{\label{fig:spectra} Spectra of $\pi^0$, $p$, $K_0^s$ and $\Lambda+\bar
  \Lambda$ as a function of $P_T$ at midrapidity in central Au+Au collisions
  at $\sqrt{s}=200$ GeV \cite{FMNB:03prc}. Dashed lines are hadrons from
  recombination of the thermal phase, dotted line is pQCD with energy loss,
  solid line is the sum of both contributions. Data are from PHENIX ($\pi^0$, 
  $p$) \cite{PHENIX:03pi0,PHENIX:03ppi} and STAR ($K_0^s$, 
  $\Lambda+\bar\Lambda$) \cite{STAR:03llbar}.}
\end{center}
\end{figure}

A very good description of hadron spectra and hadron ratios measured at
RHIC can be achieved by combining hadron production from recombination 
for intermediate transverse momentum with a perturbative calculation using
fragmentation and energy loss in the medium \cite{FMNB:03prl,FMNB:03prc}.
Fig.\ \ref{fig:spectra} shows the $P_T$ spectrum of $\pi^0$, $p$, $K_0^s$ and 
$\Lambda+\bar\Lambda$ in central Au+Au collisions obtained in 
\cite{FMNB:03prc}. The agreement with data is very good 
for $P_T > 2$ GeV/$c$. We note that the hadron spectra exhibit an 
exponential shape up to about 4 GeV/$c$ for mesons and up to about 
6 GeV/$c$ for baryons, where recombination of thermal quarks dominates. 
Above, the spectra follow a power-law and production is dominated by 
fragmentation.


\begin{figure}
\begin{center}
\epsfig{file=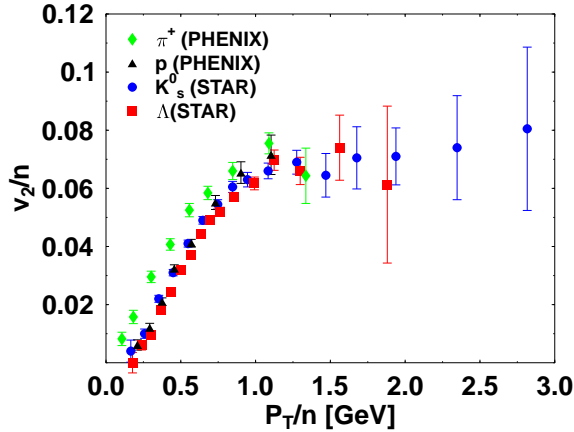,width=8cm}
\caption{\label{fig:v2} Elliptic flow $v_2$ for $\pi^+$, $p$, $K_0^s$ and
$\Lambda$ as a function of $P_T$ scaled by the number of valence quarks $n$ 
vs $P_T/n$. The data follows a universal curve, impressively confirming the 
quark scaling law predicted be recombination. Deviations for the pions are 
discussed in the text. Data are taken from PHENIX 
($\pi^+$, $p$) \cite{PHENIX:03v2} and STAR ($K_0^s$, $\Lambda$) 
\cite{STAR:03v2}.}
\end{center}
\end{figure}

Let us now assume the parton phase exhibits elliptic flow 
$v_2^{\mathrm p}(p_T)$. 
Recombination makes a prediction for elliptic flow of any hadron species
after recombination \cite{Voloshin:02,FMNB:03prc}:
\begin{equation}
  \label{eq:v2}
  v_2(P_T) = n v_2^{\mathrm p}(P_T/n).
\end{equation}
Here $n$ is the number of valence quarks for the hadron. 
Note that this scaling law is derived using the assumption of infinitely 
narrow wave functions, such that the momentum is shared equally between 
the valence partons.
Fig.\ \ref{fig:v2} shows the measured elliptic flow $v_2$ for several
hadron species in a plot with scaled axes $v_2/n$ vs $P_T/n$. All data
points (with exception of the pions) fall on one universal curve.
This is a an impressive confirmation of the quark scaling rule and the 
recombination model. The pions are shifted to lower $P_T$, 
because most pions in the detectors, even at intermediate $P_T$,
are not from hadronization, but from secondary decays of hadrons. 
This together with allowing a finite width of the wave function can improve 
the description of the pion data \cite{GreKo:04rho}. 

The quark scaling of elliptic flow shows that the relevant degrees of freedom 
at early times in the collision are partons and they prove that these
partons behave collectively.

\section{More on Elliptic Flow}

The $\phi$ meson has long been discussed to provide a good test for the 
validity of
the recombination model \cite{NMABF:03}. $\phi$ mesons are as heavy as 
protons and the questions is whether they follow the pattern
of the other much lighter mesons, or whether they behave like protons and
Lambdas.
Data from RHIC now impressively confirm that the elliptic flow
and nuclear modification factors of the $\phi$ are very similar to those 
for kaons \cite{PalCai:05phi}. This is another success for the 
recombination model.

Nevertheless one should ask to which accuracy one expects the scaling
law for elliptic flow to hold. In particular, are the scaling factors
of 2 and 3 indeed excluding any higher Fock states in the hadrons?
A recent study found that higher Fock states in an expansion
\begin{equation}
  |p\rangle = a_0 |uud \rangle + a_1 | uudg \rangle + 
  \ldots
\end{equation}
could actually be accommodated \cite{MFB:05}. 
It is easy to check that the hadron yields from a thermal
parton spectrum do not change if additional partons are allowed to coalesce.
Generally speaking the probability to form a cluster on $n$ particles
with fixed momentum $P$ from a thermal bath is independent of $n$.

\begin{figure}
\begin{center}
\epsfig{file=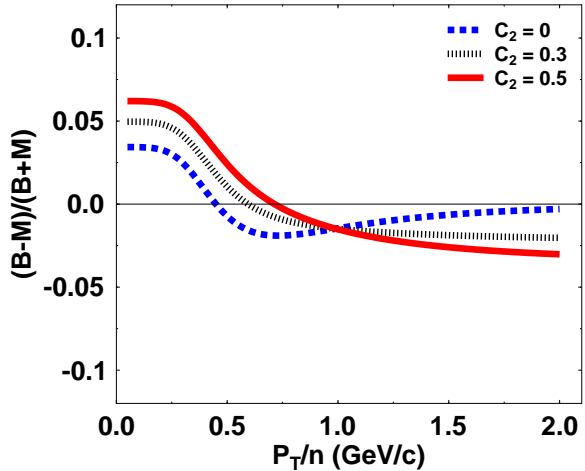,width=8cm}
\caption{\label{fig:hf} Scaling violation variable $A=(B-M)/(B+M)$ for
three different probabilities $C_2$ to have an additional parton beyond
the valence quark structure. Wave functions of finite width have been
used which lead to a scaling violation even in the case $C_2 = 0$ of 
pure valence quark recombination \cite{MFB:05}.}
\end{center}
\end{figure}

The situation changes for elliptic flow. Higher Fock states with $n$
partons come with their own scaling factor $n$ which seems to destroy 
the scaling with the number of valence quarks. However, under the assumption
that the lowest Fock state is still dominating, the numerical effect of
the corrections is surprisingly small. Fig.\ \ref{fig:hf} shows the
expected violation of the scaling law using the new asymmetry variable
$A=(B-M)/(B+M)$ where $B$ and $M$ are the {\it scaled} elliptic flow of a 
meson and a baryon respectively \cite{MFB:05}. One curve shows $A$ for 
the lowest 
Fock state only and the other two correspond to hadrons which have 
a higher Fock state component with one additional parton with probability 30\% 
and 50\% respectively. Realistic wave functions with finite width
have been used which leads to a scaling violation even for a pure
valence quark configuration. Generally the violations are smaller than
5\%. New data from STAR analyzes scaling violations in the data and
finds them to have the predicted sign and order of magnitude 
\cite{Sorensen:05}.

The study did not specify the exact nature of the additional partons.
They could be quark antiquark pairs or gluons. It was discussed 
for a long time what the fate of gluons in the recombination model is.
One can argue that they have to dress the quarks just above the critical
temperature. After this study it is clear that there is some room to 
accommodate gluons during the recombination process directly.
Further investigations in this direction are necessary.

\section{Hadron Correlations}

More and more data on dihadron correlations are available from the RHIC
experiments \cite{STARcorr:02,phenix:04corr}. 
The full picture shown by the data seems to be complicated and rich in detail.
Before one can even begin to analyze the full data set there is one 
obvious question.
The measurements find jet-like correlations at intermediate transverse
momenta that seem to be coming from jet fragmentation rather than
recombination. How can this be reconciled with the conclusion from
single hadron measurements that recombination is the dominant source of
hadrons in this kinematic regime?

Originally recombination was successfully applied to describe hadron spectra
and elliptic flow starting from assumptions about the parton phase at
hadronization. One crucial simplification always implemented
is a factorization of any $n$-parton Wigner function into a product
of independent single parton distributions
\begin{equation}
  W_{1,\ldots,n} = \prod_{i=1}^n w_i
  \label{eq:prod}
\end{equation}
By definition this factorization does not permit any correlations between 
partons. Consequently, no hadron correlations can emerge via recombination.
It has to be emphasized that the above factorization was chosen for 
simplicity and it was justified because single inclusive hadron spectra 
could be described very well.

It has been shown in \cite{FMB:04} that modifications of (\ref{eq:prod}) 
including correlations between partons indeed lead to correlations between 
hadrons after recombination. The quality of the description of single 
hadron spectra does not suffer in the process.
The source of jet-like correlations in the parton medium is the strong
coupling of jets to the medium.
The energy loss is estimated to be up to 14 GeV/fm for a 10 GeV parton 
\cite{Wang:04jq}. This implies that most jets apart from those close to 
the surface are completely stopped, dumping their energy and momentum 
into a cell of about 1 fm$^3$ in the rest frame of the medium. This results 
in a dramatic local heating, creating a hot spot in the fireball.
Moreover, the directional information of the jet is preserved. 
Partons of such a hot spot exhibit jet-like correlations. 

In \cite{FMB:04} a simple extension of the correlation-free factorization
(\ref{eq:prod}) was considered (see also \cite{FB:05}). It is assumed that 
correlations are a small 
effect and that one can restrict them to 2-particle correlations $C_{ij}$.
Then a 4-parton Wigner function can be written
\begin{equation}
  W_{1234} \approx w_1 w_2 w_3 w_4 \big( 1 + \sum_{i<j}C_{ij} \big).
  \label{eq:newfact}	
\end{equation}
The correlation functions $C_{ij}$ between parton $i$ and parton $j$ can 
be arbitrary, but one assumes that they vary slowly with momentum
and that they are only non-vanishing in a subvolume $V_c$ of the fireball.
A Gaussian ansatz $C_{ij} \sim c_0 e^{-(\phi_i-\phi_j)^2/(2 \phi_0^2)}$
seems to be reasonable to describe correlations in azimuthal angle.
The 2-meson yield is given by a convolution of the partonic Wigner function
$W_{1234}$ with the Wigner functions $\Phi_A$, $\Phi_B$ of the mesons with
an additional integration over the hadronization hypersurface $\Sigma$
\cite{FMB:04}. It is assumed that the correlation strength $c_0 \ll 1$ which 
permits omitting quadratic terms like $c_0^2$ or $c_0 v_2$.

We can now study the associated yield $Y_{AB}$ for a given trigger hadron
$A$ in a given kinematic window as a function of the relative azimuthal
angle $\Delta \phi$ between the two. One finds
\begin{equation}
  2\pi N_A Y_{AB} (\Delta\Phi) = Q \hat c_0 e^{-(\Delta\Phi)^2/(2\phi_0)^2}
  N_A N_B.
  \label{eq:res}
\end{equation}
The $N_i$ are single particle yields in the kinematic window of the trigger 
meson or associated meson and $\hat c_0 = c_0 V_c/(\tau A_T)$ where
$\tau A_T$ is the hadronization volume.
The factor $Q=4$ (for two mesons) indicates an enhancement of the correlations 
in the hadron phase compared to the parton phase. The effect is the same as for
the amplification of elliptic flow by the number $n$ of valence quarks in the
hadron. In the case of 2-parton correlations, $Q$ counts the number of 
possible correlated pairs between the $n_A$ (anti)quarks of meson $A$ and the
$n_B$ (anti)quarks of meson $B$. In the weak correlation limit 
where quadratic terms are suppressed only single correlations are counted.
Apparently one has
\begin{equation}
  Q=n_A n_B,
\end{equation}
thus $Q=6$ for a meson-baryon pair and $Q=9$ for a baryon-baryon pair.

\begin{figure}
\begin{center}
\epsfig{file=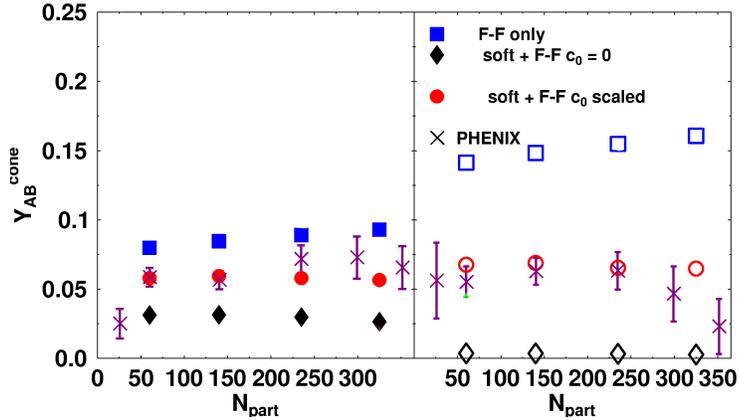,width=10cm} \\
\caption{\label{fig:res} The associated yield $Y_{AB}^{\mathrm cone}$ for
baryon triggers (right panel) and meson triggers (left panel) as a function
of $N_{\mathrm part}$. Squares: fragmentation only; diamonds: fragmentation
and recombination with $\hat c_0=0$;
circles: the same with $\hat c_0=0.08\times 100/N_{\mathrm part}$;
data: PHENIX \cite{phenix:04corr}.}
\end{center}
\end{figure}

Fig.\ \ref{fig:res} shows the associated yield of hadrons integrated 
in azimuthal angle around the near side ($\Delta\phi =0$)
for the case that the trigger is a baryon (proton or antiproton) and 
a meson (pion or kaon) for different centralities. 
The kinematic window is 2.5 GeV/$c \le P_{TA}\le$4.0 GeV/$c$ for trigger 
particles and 1.7 GeV/$c\le P_{TB}\le$2.5 GeV/$c$ for associated particles, 
and $|y_A|$, $|y_B|<0.35$.
Fig.\ \ref{fig:res} shows the associated yield with only fragmentation,
and fragmentation and recombination both taken into account together
with PHENIX data \cite{phenix:04corr}. A good description of the data can
be reached assuming a constant correlation volume.
The parameters used for the fireball are the same that lead to 
a good description of single hadron spectra and elliptic flow
\cite{FMNB:03prc}.

\section{Soft-Hard Recombination}

Some implementations of recombination take into account the possibility
that hard partons can recombine with soft partons 
\cite{GreKoLe:03prl,HwaYa:03}. This soft-hard recombination smoothens
the transition between the pure fragmentation and pure (thermal)
recombination regions. It was predicted that typical recombination features
like a larger baryon/meson ratio would then extend to even higher $P_T$,
beyond 6 GeV/$c$. Indeed such behavior might have been seen in recent data 
\cite{star:06id}.

Soft-hard recombination could be a correction to fragmentation which
is even important in much more dilute systems. The reason is that
fragmentation is an extremely ineffective mechanism to create baryons.
Soft-hard recombination is a very good candidate to explain the very larger
Cronin enhancement for baryons in $d+$Au collisions \cite{HwaYa:04da}.
There is no thermal parton phase created in $d+$Au collisions, but
the existence of a cloud of soft partons with exponential spectrum is
sufficient to boost baryon production via coalescence.
Soft-hard recombination could even be responsible for the suppression of
hadrons at forward rapidities in $d+$Au \cite{HYF:04da}.

\ack I would like to thank B.\ M\"uller, S.\ A.\ Bass and C.\ Nonaka.
This work was supported by DOE grant DE-FG02-87ER40328.

\end{document}